\documentclass[11pt]{article}
\newcommand{\Comment}[1]{{}}
\usepackage[textwidth = 430 pt, textheight = 630 pt]{geometry}
\usepackage{amssymb,euscript,amsmath,amsfonts}

\usepackage{color}
\definecolor{MyDarkBlue}{rgb}{0.15,0.15,0.45}
\usepackage[linktocpage=true]{hyperref}
\hypersetup{
colorlinks=true,
citecolor=MyDarkBlue,
linkcolor=MyDarkBlue,
urlcolor=MyDarkBlue,
pdfauthor={Neil Lambert },
pdftitle={ },
pdfsubject={hep-th}
}

\usepackage[numbers,sort&compress]{natbib}
\usepackage{hypernat}

\newcommand{\ret}{\nonumber \\}

\newcommand{\be}{\begin{equation}}
\newcommand{\ee}{\end{equation}}
\newcommand{\bea}{\begin{eqnarray}}
\newcommand{\eea}{\end{eqnarray}}

\begin{document}

\renewcommand{\thefootnote}{\fnsymbol{footnote}}

   \vspace{1.8truecm}

 \centerline{\LARGE \bf {\sc   $(2,0)$ Lagrangian   Structures }}

\centerline{\LARGE \bf {\sc  }} \vspace{2truecm} \thispagestyle{empty} \centerline{
    {\large {\bf {\sc Neil~Lambert${}^{\,a,}$}}}\footnote{E-mail address: \href{mailto:neil.lambert@kcl.ac.uk}{\tt neil.lambert@kcl.ac.uk}} 
  }

\vspace{1cm}
\centerline{${}^a${\it Department of Mathematics}}
\centerline{{\it King's College London }} 
\centerline{{\it  WC2R 2LS, UK}} 

\vspace{1.0truecm}

 
\thispagestyle{empty}

\centerline{\sc Abstract}
\vspace{0.4truecm}
\begin{center}
\begin{minipage}[c]{360pt}{
    \noindent}
By including an additional  self-dual three-form we construct a Lorentz invariant lagrangian for the abelian $(2,0)$ tensor supermultiplet. The extra three-form is a supersymmetry singlet and  decouples from the $(2,0)$ tensor supermultiplet. We also present an interacting non-abelian generalization which  reproduces the equations of motion of \cite{Lambert:2010wm} and can describe  some aspects of two interacting M5-branes. 
\end{minipage}
\end{center}

\newpage 
 \section{Introduction}\label{sect: Intro}
 
There are good reasons to believe that there is no lagrangian   formulation for  the six-dimensional $(2,0)$-theories that describe M5-branes.  Some of the  various arguments can be summarised as follows:
 
 \begin{itemize}
 \item Reduce the $(2,0)$-Theory on a compact four-manifold ${\cal M}$. The presence of the self-dual 3-form would lead to $\sigma({\cal M}) = b_2^+({\cal M})-b_2^-({\cal M})$ chiral bosons in the resulting two-dimensional theory. If one had an action then  one would expect there to be a modular invariant partition function coming from the $SL(2,{\mathbb Z})$  of large  diffeomorphisms in the remaining two-dimensions, which we take to be a torus of finite size. However such a  partition function only exists if $\sigma$ is a multiple of $8$.  In particular one expects to be able to embed  ${\mathbb CP}^2$, which has $\sigma({\cal M})=1$, in to M-theory  and the resulting two-dimensional theory cannot have a modular invariant partition function \cite{Witten:1996hc}.

 \item Reducing the ${\mathfrak su}(2n)$ $(2,0)$-Theory on $S^1$,  along with an outer automorphism twist,  leads to five-dimensional  maximally supersymmetric Yang-Mills with gauge group  ${\mathfrak so}(2n+1)$. But ${\mathfrak so}(2n+1)$ is not a subalgebra of ${\mathfrak su}(2n)$ for generic $n$ \cite{Tachikawa:2011ch}.
 
  \item  The standard M-theory dictionary states that reducing the $(2,0)$-Theory  on $S^1$ of radius $R$ we should find five-dimensional maximally supersymmetric Yang-Mills with coupling proportional to $R$, meaning that the five-dimensional action is inversely proportional to $R$. However dimensional reduction of an action in six-dimensions   naively leads to a five-dimensional action that is directly proportional to $R$ \cite{Witten:2009at}.

\item There are no satisfactory deformations of the free Lagrangian \cite{Bekaert:1999dp}. In addition there is no sequence of interacting six-dimensional superconformal field theories that converge to a free theory \cite{Chang:2018xmx}.
  
   \item There are no interacting, power counting renormalizable, lagrangians in six-dimensions with an energy-momentum tensor that is bounded from below  \cite{textbook}. However the $(2,0)$ theory is a conformal field theory and should have no dimensionful parameters  and   be UV complete.
 
 \end{itemize}
 
  These arguments are quite convincing and hence we don't expect to find a definitive lagrangian for the interacting $(2,0)$ theory. However   it is worth exploring what lagrangian structures exist and what they can do. Furthermore by hunting for unicorns  we may find other creatures that are   useful in understanding the theory more generally. In particular there  are some possible ways out of the first two points:
\begin{itemize}
\item Rokhlin's theorem states that for any compact four-dimensional spin manifold $\sigma({\cal M}) $ is a multiple of $16$ and hence the dimensionally reduced theory can have a modular invariant partition function. The problem could be that the action must be coupled to background fields in a non-standard way so as to allow for a non-spin manifold such as ${\mathbb CP}^2$. 

\item For $n=1$ ${\mathfrak so}(3) ={\mathfrak su}(2 )$ so this objection fails. This is  reminiscent of the M2-brane story for which a lagrangian   with all the supersymmetries manifest only exists for two M2's.

\end{itemize}

This work was influenced by Sen who has introduced a method to formulate an action for self-dual abelian fields in $4n+2$ dimensions \cite{Sen:2015nph,Sen:2019qit} by including a second self-dual form which then decouples. This construction has the feature that the coupling to gravity is somewhat non-standard so that  diffeomorphisms act differently from usual and hence  provides hope that the first and third issues can be overcome, although we will not discuss this here.

Thus the  purpose of this paper is  twofold. The first is to construct a new action for the abelian  (free) $(2,0)$ multiplet. We do this by introducing an additional self-dual three form which is a supersymmetry singlet. The second  is to explore how one might generalise it to  a non-abelian theory, at least for two M5-branes, and see how far we get. In the latter case  we must  be willing to be  suitably creative. We will postpone for later the issue of whether or not the resulting dynamical theories are well-defined and how much of the $(2,0)$ theory they capture. We are more interested in exploring the possible structures with a hope that they will lead to additional insights that will be fruitful, even  without a lagrangian.
  
We would also like to mention other related work. Using the notion of tensor hierarchy a class of six-dimensional $(1,0)$ Lagrangians was obtained in \cite{Samtleben:2011fj} but the self-duality condition was imposed `by hand' on the equations of motion. The use of an additional self-dual three-form to construct actions for self-dual three-forms has appeared in the Twistor approach of  \cite{Mason:2011nw}, \cite{Saemann:2011nb} and was generalised to a non-abelian but flat gauge fields in \cite{Jurco:2019bvp}. Mathematically focused discussions of lagangian structures for the $(2,0)$-theory also recently appeared in \cite{Saemann:2017rjm,Saemann:2019dsl}. Even more recently an alternative construction of self-dual forms was given in \cite{Mkrtchyan:2019opf}.

The rest of this paper is organised as follows. In section 2 we will review the construction of Sen for the particular case of a self-dual three-form in six-dimensional Minkowski space. In section 3 we will adapt this to the case of an abelian supersymmetry $(2,0)$ multiplet, including potential external interactions. In section 4 we will examine how we might introduce an interacting 
 $(2,0)$ theory, leading to an action (or more precisely a family of  actions)  which reproduces the equation of motion of the $(2,0)$ theory of \cite{Lambert:2010wm}. Finally in section 5 we state our results and conclusions.

 \section{The $(2,0)$ Multiplet and Sen's Prescription}\label{sect: Review}
 
 The linearised equations of motion for the $(2,0)$ tensor multiplet can be written  as 
 \begin{align}
 \partial_\mu\partial^\mu X^I &= 0\ret
 i\Gamma^\mu\partial_\mu \Psi & = 0\ret
dH& =0\ ,
 \end{align}
 where $H=\star H$ is a self-dual 3-form and $\Psi$ is a chiral spinor with 8 real on-shell degrees of freedom: $\Gamma_{012345}\Psi=-\Psi$. We use conventions where  $\mu,\nu =0,1,2,3,4,5$, $I=6,7,8,9,10$, $\eta_{\mu\nu}={\rm diag}(-1,1,1,1,1,1)$,  $\varepsilon^{012345}=1$ and $(\Gamma_\mu,\Gamma^I)$ form a real representation of the $Spin(1,10)$ Clifford algebra and all spinors are real.

These equations are invariant under the on-shell $(2,0)$ supersymmetries:
\begin{align}\label{OSsusy}
\delta X^I &=i\bar\epsilon \Gamma^I\Psi\ret
\delta H_{\mu\nu\lambda} & =  {3i} \bar\epsilon\Gamma_{[\mu\nu}\partial_{\lambda]}\Psi 
\ret
\delta \Psi &= \Gamma^\mu\Gamma^I\partial_\mu X^I\epsilon+\frac{1}{2\cdot  3!}\Gamma^{\mu\nu\lambda}H_{\mu\nu\lambda}\epsilon\ ,
\end{align}
where $\Gamma_{012345}\epsilon=\epsilon$. These close, on-shell, onto translations. Alternatively one often introduces a two-form $b$ so that $H=db$ with $\delta b_{\mu\nu}= i\bar\epsilon \Gamma_{\mu\nu}\Psi$.

Let us now review the action proposed in \cite{Sen:2015nph,Sen:2019qit}:
\begin{equation}\label{A1}
S =  \int  \left[ 
 \frac{1}{4}dB\wedge\star dB -H\wedge dB
+ {\cal L}_{int}(H) + {\cal L}_m(X^I,\Psi)\right]
\end{equation}
Here we have relabelled fields so as to conform more closely to the standard $(2,0)$ literature. In particular, in the notation of \cite{Sen:2015nph,Sen:2019qit} $B=\sqrt{2}P$ and $H=- Q/ \sqrt{2}$. We have also split the interaction term ${\cal L}_I$  that appears in \cite{Sen:2015nph,Sen:2019qit}  into one that depends on $H$ and the rest ${\cal L}_m$ which includes the kinetic terms for the remaining fields to facilitate our discussion. 

We use a convention where, for a $p$-form $\omega$,
\begin{align}
\omega &= \frac{1}{p!}\omega_{\mu_1...\mu_p}dx^{\mu_1}\wedge ... dx^{\mu_p}\ret
d\omega & = \frac{1}{p!}\partial_{\nu}\omega_{\mu_1...\mu_p}dx^\nu\wedge dx^{\mu_1}\wedge ... dx^{\mu_p}\ .
\end{align}
The Hodge dual acts on $p$-form components as
\begin{equation}
(\star \omega)_{\mu_1..\mu_{6-p}} = \frac{1}{p!}\varepsilon_{\mu_1...\mu_{6-p}\nu_{1}...\nu_p}\omega^{\nu_1...\nu_p}\ .
\end{equation}
This satisfies $\star^2 = 1$ on odd forms, $\star^2 = -1$ on even forms and $ \omega\wedge \star \chi = \chi\wedge \star \omega$ for two $p$-forms.

Next we observe that the equations of motion for $B$ and $H$ that result from this action can be written as
\begin{align}\label{OS}
d\left(\frac{1}{2}(dB+\star dB) + H\right) &= 0 \ret
 -\frac{1}{2}(dB-\star dB) - R   & = 0\ ,
\end{align}
where the anti-self-dual 3-form $R$ is defined by
\begin{equation}
\delta {\cal L}_{int} = -\int\delta H\wedge R\ . 
\end{equation}
Note that  $R=-\star R$ by construction\footnote{We have also rescaled $R\to -2R$ compared to \cite{Sen:2015nph,Sen:2019qit}.}.
Thus we see that there are two self-dual 3-forms $\tfrac12(dB+\star dB) $ and $H$. The first one has the wrong sign kinetic term but, as shown in \cite{Sen:2019qit}, the combination   $\tfrac12(dB+\star dB) + H$ is free (closed) and decouples. This is most transparently seen in the Hamiltonian formulation. Thus it can be safely discarded from any physical quantities.  The physically relevant 3-form  is $H$ which is not closed but rather has a source: 
\begin{equation}
\star d\star H  = J\qquad\Leftrightarrow\qquad dH = -\star J \ ,
\end{equation}
where $ J = -\star dR$ is a 2-form current. Of course these two equations are equivalent and can also be written more succinctly as $d(H+ R)=0$ but the above form seems more suggestive. Note that since $R=-\star R$ one can't simply solve this  by taking $H=-R + dC$ and imposing $dC=\star dC$.

\section{An Abelian (2,0) Action}\label{sect: SUSY}

Our first task is to extend the action (\ref{A1}) to the free $(2,0)$ multiplet by setting ${\cal L}_{int}=0$ and choosing a suitable ${\cal L}_m$. Thus we consider 
\begin{equation}\label{Sgeneral}
S = \int  \ \left[
\eta dB\wedge\star dB -H\wedge dB
- \frac12 \partial_\mu X^I\partial^\mu X^I  + \frac {i}{2}\bar\Psi\Gamma^\mu\partial_\mu \Psi\right]\ ,
\end{equation}
where $\eta$ is a constant to be determined. In particular the usual sign kinetic term for $B$ requires $\eta<0$. We postpone interaction terms to the next section. Here we wish to establish supersymmetry of this free action. To this end we consider the ansatz:
\begin{align}\label{susy}
\delta X^I &=i\bar\epsilon \Gamma^I\Psi\ret
\delta B_{\mu\nu} & = i\beta\bar\epsilon\Gamma_{\mu\nu}\Psi\ret
\delta H_{\mu\nu\lambda} &  =   \frac{i\alpha}{2} \bar\epsilon\Gamma_{[\mu\nu}\partial_{\lambda]}\Psi + \frac{i\alpha}{2\cdot 3!}\varepsilon_{\mu\nu\lambda\rho\sigma\tau}\bar\epsilon\Gamma^{\rho\sigma}\partial^{\tau}\Psi \ret
\delta \Psi &= \Gamma^\mu\Gamma^I\partial_\mu X^I\epsilon+\frac{\gamma}{3!}\Gamma^{\mu\nu\lambda}H_{\mu\nu\lambda}\epsilon+ \frac{\delta}{2!}\Gamma^{\mu\nu\lambda}\partial_\mu B_{\nu\lambda}\epsilon \ .
\end{align}
 Note that we have to modify the on-shell transformation for $H$ that was in (\ref{OSsusy}) to ensure that $\delta H$ is self-dual off-shell.

We find the action  (\ref{Sgeneral}) is invariant if
\begin{equation}\label{C1}
 \gamma = -\frac{\beta}{2}\qquad \delta = -2\eta\beta-\frac{\alpha}{6} \ .
\end{equation}
Furthermore closure on $X^I$ leads to the translation
\begin{equation}
[\delta_1,\delta_2]X^I = -2i\bar\epsilon_2\Gamma^\mu\epsilon_1 \partial_\mu X^I\ .
\end{equation}
On the other hand on-shell closure on the fermions requires that
\begin{equation}\label{C2}
\frac{2}{3}\alpha\gamma + 2\beta\delta = 1 \ .
\end{equation}

However we have a reducible representation of supersymmetry as there are  two self-dual 3-forms: $dB+\star dB$ and $H$.  In particular we observe  that
\begin{equation}
H_{(s)} =  \frac{1}{2}\left(dB  +\star dB\right)  -\frac{3\beta }{\alpha}H   
\end{equation}
satisfies $\delta H_{(s)}=0$ and hence is a supersymmetry singlet (of course if $\alpha=0$ then $H$ is the supersymmetry singlet). We also note that for the general form of the action (\ref{Sgeneral}) the free  combination is
\begin{equation}
H_{(f)} =  \frac{1}{2}\left(dB  +\star dB\right)   +\frac{1}{ 4\eta}H \ .
\end{equation}
Therefore we choose\begin{equation}
\beta = -\frac{\alpha}{12\eta}\ ,
\end{equation} 
so that $H_{(f)} = H_{(s)}$  (if $\alpha=0$ then we would take $\eta=0$ and $H_{free}=H_{(s)}=H$). This in turn implies   
$\alpha^2 = 36\eta$ so that $\eta>0$ and hence $B$ must have the wrong sign kinetic term. 

The action (\ref{Sgeneral}) has the peculiar symmetry
\begin{equation}\label{vSym}
\tilde\delta  B = i_v H \qquad \tilde\delta  H = -2\eta (d\tilde\delta  B + \star d \tilde\delta  B)\ ,
\end{equation}
where $(i_v H)_{\mu\nu} = v^\lambda H_{\mu\nu\lambda}$ and $v^\lambda$ is 
 any constant vector. Note that  $\tilde\delta  H_{(s)}=0$. We can now evaluate the  closure on $B$ and  $H$ to find
\begin{align}
[\delta_1,\delta_2] B_{\mu\nu}  & = \partial_{[\mu}\left(4i \bar\epsilon_2\Gamma^I\Gamma_{\nu]}X^I\epsilon_1 \right)  + \tilde\delta B_{\mu\nu}\nonumber\\
[\delta_1,\delta_2] H_{\mu\nu\lambda} &= \tilde\delta H_{\mu\nu\lambda}\ .
\end{align}
where $v^\rho=-2\beta^2i\bar\epsilon_2\Gamma^\rho\epsilon_1$ and the first term in $[\delta_1,\delta_2] B_{\mu\nu}$ is a gauge transformation. On-shell, {\it i.e.} for $dH=0$, we have
\begin{equation}
(\tilde\delta  H)_{\mu\nu\lambda} = v^\rho\partial_\rho H_{\mu\nu\lambda}  \ ,
\end{equation}
so that the supersymmetries close onto a translation of $H$ and to find the same transformation that we did for $X^I$ and $\Psi$ requires $\beta=\pm 1$. For $B$ we can re-write the closure as 
\begin{align}
[\delta_1,\delta_2] B_{\mu\nu}  & = -2i\bar\epsilon_2\Gamma^\lambda\epsilon_1 \partial_\lambda B_{\mu\nu}+ 2\partial_{[\mu}\left(2i \bar\epsilon_2\Gamma^I\Gamma_{\nu]}X^I\epsilon_1 -2i\bar\epsilon_2\Gamma^\lambda\epsilon_1 B_{\nu]\lambda}\right)\nonumber\\
&
+2i\bar\epsilon_2\Gamma^\lambda\epsilon_1 \left(H_{(s)}+\frac{1}{2}(dB-\star dB)\right)_{\mu\nu\lambda}\ .
\end{align}
 Here the first term is a translation, the second a gauge transformation and second line, on-shell, is just $H_{(s)}$.


For concreteness and to agree with the conventions in section \ref{sect: Review} we take (changing the sign of $\beta$  merely changes the signs of $\alpha$ and $\gamma$)  \begin{equation}
\eta=1/4\qquad \alpha = 3  \qquad \beta = -1\qquad \gamma =  \frac{1}{2 }\qquad  \delta=0 \ .
\end{equation}
In summary the action is
\begin{equation}\label{Sis}
S = \int  \ \left[
\frac14 dB\wedge\star dB -H\wedge dB
- \frac12 \partial_\mu X^I\partial^\mu X^I  + \frac {i}{2}\bar\Psi\Gamma^\mu\partial_\mu \Psi\right]\ ,
\end{equation}
and this is invariant under the supersymmetry
\begin{align}\label{susy}
\delta X^I &=i\bar\epsilon \Gamma^I\Psi\ret
\delta B_{\mu\nu} & = -i\bar\epsilon\Gamma_{\mu\nu}\Psi\ret
\delta H_{\mu\nu\lambda} &  =   \frac{3i }{{2}} \bar\epsilon\Gamma_{[\mu\nu}\partial_{\lambda]}\Psi + \frac{3i }{{2}\cdot 3!}\varepsilon_{\mu\nu\lambda\rho\sigma\tau}\bar\epsilon\Gamma^{\rho\sigma}\partial^{\tau}\Psi - \frac{i}{2}\partial^\rho\bar\epsilon\Gamma_\rho\Gamma_{\mu\nu\lambda}\Psi\ret
\delta \Psi &= \Gamma^\mu\Gamma^I\partial_\mu X^I\epsilon+\frac{1}{2\cdot 3!}\Gamma^{\mu\nu\lambda}H_{\mu\nu\lambda}\epsilon -\frac{2}{3}\Gamma^IX^I\Gamma^\rho\partial_\rho\epsilon\ .
\end{align}
where we have also extended our results to allow for superconformal symmetries with
\begin{equation}
\partial_\mu \epsilon = \frac{1}{6}\Gamma_\mu\Gamma^\rho\partial_\rho\epsilon\ .
\end{equation}
Note that the last term in $\delta H_{\mu\nu\lambda}$ is self-dual. For constant $\epsilon$ and on-shell fermions this agrees with the abelian supersymmetries in \cite{Lambert:2010wm} (except that $B_{\mu\nu}$, and hence $\delta B_{\mu\nu}$, does not appear there).


\subsection{External Interactions}\label{sect: Interactions}

Next we want to see if we can  introduce an interaction term into the action while preserving supersymmetry. In this section we will restrict to cases where the interactions arise from external sources and not from the fields in the $(2,0)$ multiplet. To begin with we take
\begin{equation}
S =  \int \ \left[
\frac{1}{4} dB\wedge\star dB - H\wedge dB 
- \frac12 \partial_\mu X^I\partial^\mu X^I + \frac {i}{2}\bar\Psi\Gamma^\mu\partial_\mu \Psi+ {\cal L}_{int}(H)\right]\ ,
\end{equation}
where ${\cal L}_{int}(H)$ depends on $H$ but not $X^I$, $\Psi$ or $B$. Such terms appear in   \cite{Sen:2015nph,Sen:2019qit} as the coupling of $H$ to the metric and external sources.

Taking its variation under supersymmetry we find
\begin{equation}
\delta {\cal L}_{int} = -\delta H\wedge R =    \bar \partial_\rho\Psi\Gamma^{\mu\nu\lambda\rho} \epsilon  R_{\mu\nu\lambda} \cong     -\bar\Psi \partial_\rho R_{\mu\nu\lambda}  \Gamma^{\rho\mu\nu\lambda}\epsilon  \ ,
\end{equation}
where we have used the fact that $R=-\star R$. Clearly this will be invariant if $dR=0$ and this will also preserve the symmetry $\tilde\delta $. However $dR=0$ also means that the source $J=0$. 
 To proceed we assume  that $d\star dR=0$ so that  we  can write 
\begin{equation}
dR = \star  dj\ ,\label{Rj}
\end{equation}
for some $j$. Thus in the notation of section \ref{sect: Review} we have $J=-\frac12 dj$. We further assume that $j$ can be chosen to satisfy
\begin{equation}
d\star j =0\ .\label{j}
\end{equation}
This second condition can be viewed as a sort of Lorentz gauge choice of $j$: $\partial^\mu j_\mu=0$. Or alternatively that $j_\mu$ can be thought of as a traditional 1-form conserved current. Note that this condition implies that we can write the $H$ equation as:
\begin{align}
d   H & = \star dj \ .
\end{align}
In this case we have
\begin{equation}
\delta {\cal L}_{int} \cong   -6i\bar\Psi \Gamma^\mu \Gamma^\nu \partial_\mu  j_\nu \epsilon  \ ,
\end{equation}
which means that we can restore supersymmetry of the action by replacing
\begin{equation}
\delta \Psi \to \delta\Psi  +6 \Gamma^\nu j_\nu \epsilon\ .
\end{equation}


\subsection{Comments On Alternative Prescriptions}\label{Alt}

For educational purposes let us consider alternative ways to introduce an interaction term ${\cal L}_{int}$ that depends on $B$ instead of  $H$. In this case we write
\begin{equation}
\delta {\cal L}_{int} =  -\delta B\wedge \star T\ .  
\end{equation}
If we look at the supersymmetry then we find
\begin{equation}
\delta {\cal L}_{int}  = 12i \bar\Psi \Gamma^{\mu\nu}\epsilon T_{\mu\nu}\ .
\end{equation}
To continue we  assume that
\begin{equation}
T = d k \ ,
\end{equation}
with
\begin{equation}
d\star k =0 \ .
\end{equation}
We now see that 
\begin{equation}
\delta {\cal L}_{int} \cong 12i\bar\Psi \Gamma^{\mu}\Gamma^{\nu}\partial_\mu k_\nu\epsilon\ .
\end{equation}
 and again we can cure this by the replacement 
 \begin{equation}
\delta \Psi \to \delta\Psi  -  24  \Gamma^\nu k_\nu \epsilon\ .
\end{equation}

In this case the equations of motion for $H$ and $B$ lead to
\begin{align}
d   H  &=   \star T \ret
dB-\star dB  & = 0\ .
\end{align}
Here the natural self-dual supersymmetry singlet is $dB$ but this is not consistent with supersymmetry of the action since $\beta=0$ cannot solve the constraints (\ref{C1}) and (\ref{C2}). We could take $H$ to be the supersymmetry singlet, so $\alpha=0$ (and we take $\beta=1$,\ $\gamma=-1/2$,\ $\eta=-1/4$), but then we simply have an interacting supersymmetry singlet $H$ along with with a free $(2,0)$ multiplet $(B,X^I,\Psi)$.

Lastly we can also  consider a linear combination of the two interaction terms. In particular if we had a sources which satisfy $R=-\star R$ and also  $T =   \star dR$   then the action remains supersymmetric as
\begin{align}
\delta {\cal L}_{int} &= - \delta H\wedge R +\delta B\wedge dR\nonumber\\
& \cong  -\delta H\wedge R  - d\delta B \wedge R\nonumber\\
& =   -\delta H_{(s)}\wedge R\nonumber\\
& =0\ .
\end{align}
In this case we find the equations of motion are
\begin{align}\label{OS2}
d\left(\frac{1}{2}(dB+\star dB) + H- R\right) &= 0 \ret
-\frac{1}{2}(dB-\star dB)   -R   & = 0\ .
  \end{align}
Here  $dH=0$ but $H_{(s)} $ has a source
\begin{equation}
dH_{(s)} = \star J\ ,
\end{equation}
where now $J=-\star dR$. But again this is not very interesting as we simply have an interacting self-dual three-form $H_{(s)}$, which is invariant under supersymmetry, along with a decoupled free $(2,0)$ multiplet $(H,X^I,\Psi)$.



\section{Non-Abelian Extensions}

 \subsection{Flat Gauging}\label{sect: Flat}
 
It is possible to include gauge fields into the above action so long as their equation of motion sets them to be flat. In this way we do not introduce any  new local degrees of freedom. To this end we assume that each of the fields above take  values in a real vector space ${\cal V}$ with positive definite inner-product $\langle\cdot,\cdot\rangle$ and basis $T^a$, $a=1,..,N$. We then introduce a covariant derivative $D_\mu X^I = \partial_\mu X^I  - \tilde A _\mu(X^I)$ which in component form is
\begin{equation}
D_\mu X^I_a = \partial_\mu X^I_a  - \tilde A^{ r}_\mu(\tilde{T_r})_a{}^b X^I_b
\end{equation}
where $(\tilde{T_r})_a{}^b$, $a,b=1,...,N$, form an $N$-dimensional representation  of a Lie-algebra ${\cal G}$ with $r=1,...,{\rm dim}({\cal G})$.  We have used a tilde to denote the fact that the fields take values in the Lie-algebra ${\cal G}$ that  acts on the vector space ${\cal V}$ where the other fields live. We also assume that there is an invariant inner-product on ${\cal G}$ which we denote by $(\cdot ,\cdot )$.

The action is now
\begin{align}
S = \int d^6x\Big[ &\frac14\langle D B  \wedge  \star D B\rangle
  -\langle H  \wedge  D B\rangle- \frac12 \langle  D _\mu X^I   D ^\mu X^I \rangle + \frac {i}{2}\langle  \bar\Psi  \Gamma^\mu D_\mu \Psi\rangle \nonumber\\
&  +(\tilde F \wedge \tilde  W)+{\cal L}_{int}(H)\Big]\ ,
\end{align}
where 
\begin{equation}
\tilde F  = -[D,D] = d\tilde A   - \tilde A \wedge \tilde A \ .
\end{equation}
We have introduced a Lagrange multiplier  4-form $\tilde W$ that  takes values in  ${\cal G}$ and which  ensures that $\tilde A $ is a flat connection. 
Note that there is a gauge symmetry $\tilde W\to \tilde W +D \tilde \Lambda$.  

This action is supersymmetric if we simply replace $\partial_\mu \to D _\mu$ in (\ref{susy}) and furthermore take
\begin{align}\label{FlatASusy}
\delta \tilde A_\mu &=0 \ret
\delta \tilde W_{\mu\nu\lambda\rho}(\ \cdot\ ) & =   {3 i} \bar\epsilon\Gamma_{[\mu\nu}[B_{\lambda\rho] },\Psi, \ \cdot\ ] +i\bar\epsilon\Gamma_{\mu\nu\lambda\rho}\Gamma^I[X^I, \Psi,\ \cdot\ ] \ .
\end{align} 
Here we have introduced a three-algebra structure  on $\cal V$ which is a tri-linear map $[\cdot,\cdot,\cdot]:{\cal V}\otimes {\cal V}\otimes{\cal V} \to {\cal V}$ that is compatible with the gauge symmetry. To obtain such a structure one starts by constructing maps
\begin{equation}
\tilde \varphi:{\cal V}\times {\cal V} \to {\cal G}\ ,
\end{equation}
given by $\tilde \varphi(U,V)  = \sum_r \langle U,\tilde{T^r}(V)\rangle\tilde{T_r}  $ where we have used the inner-product on ${\cal G}$ to raise the $r$-index on the generators $\tilde T_r$. This in turn allows us to define a triple product on ${\cal V}$ as\footnote{Note that we do not assume here any symmetry properties of $[\cdot,\cdot,\cdot]$}
\begin{equation}
[X,Y,Z] = \tilde \varphi(X,Y)(Z)  = \sum_r \langle X,\tilde{T^r}(Y)\rangle\tilde{T_r}( Z)= f^{cdb}{}_a  X_cY_d Z_b T^a\ ,
\end{equation}
where $f^{cdb}{}_a= \sum_r \tilde{T^r}^{cd}\tilde{T_r}_a{}^b$. The compatibility condition means that we assume 
\begin{equation}
(\tilde T, [U,V,\ \cdot\ ]) =\langle \tilde T(U), V\rangle =- \langle U, \tilde T(V)\rangle \ .
\end{equation} 
We will use this relation repeated in what follows.
As a result of the of the Jacobi identity the  triple product satisfies the fundamental identity
\begin{equation}
[U,V,[X,Y,Z]] = [[U,V,X],Y,Z] +[X,[U,V,Y],Z] +[X,Y,[U,V,Z]] \ . 
\end{equation}
For  a positive definite innerproduct $\langle\ \cdot  ,\ \cdot \rangle$ there is a unique example of an irreducible  finite-dimensional three-algebra where  $[\cdot,\cdot,\cdot]$ is a totally anti-symmetric \cite{Papadopoulos:2008sk,Gauntlett:2008uf}.  In particular ${\cal V}={\mathbb R}^4$ and the associated Lie-algebra is ${\mathfrak su(2)}\oplus {\mathfrak su(2)}$.

\subsection{An Interacting Non-abelian Action}\label{sect: Nonabelian}

Next we want to consider the case where we have non-abelian interactions between the fields of the $(2,0)$ tensor multiplet. It is easy to see that there are no choices for ${\cal L}_{int}$ that depend on $H$, $X^I$ and $\Psi$ without introducing coupling constants with negative mass-dimensions and which are therefore, at least naively, non-renormalizable. Another  problem is that the condition $d^2=0$ featured heavily in the abelian analysis above but in a non-abelian theory $D^2\sim {\tilde F} \ne 0$. So we proceed we must indulge ourselves in some form of shady speculation.

In this section we follow the route explored in \cite{Lambert:2010wm} which presents an interacting  system of equations of motions for a set of fields $(H,X^I,\Psi,\tilde A_\mu)$ that generalises the free equations of motion constructed above and which are invariant under  $(2,0)$ supersymmetry. So here we wish to see if we can construct a lagrangian for this system along the lines outlined above.  

In order to construct interactions the $(2,0)$ system in \cite{Lambert:2010wm} introduces a non-dynamical vector field $Y^\mu$\footnote{$Y^\mu$ was denoted by $C^\mu$ in \cite{Lambert:2010wm}} with scaling dimension $-1$ which takes values in ${\cal V}$ and satisfies the constraints 
\begin{equation}\label{constraints}
D_\mu Y^\nu=0\qquad [Y^\mu, D_\mu (\ \cdot \ ),\ \cdot'\ ]=0\qquad   [Y^\mu,Y^\nu,\ \cdot \ ]=0 \ .
\end{equation}
Here the three-algebra is totally anti-symmetric and so we take  ${\cal V}={\mathbb R}^4$  leading to the gauge algebra ${\mathfrak su(2)}\oplus {\mathfrak su(2)}$. The second condition asserts that the non-abelian part of the theory is restricted to only depend on  five of the coordinates orthogonal to $Y^\mu$. 
The equations of motion are
 \begin{align}\label{eomfixed}
  0&=D^2 X^I-\frac{i}{2}[Y^\sigma,\bar \Psi,\Gamma_\sigma\Gamma^I\Psi]+[Y^\sigma,X^J,[Y_\sigma,X^J,X^I]]\nonumber\\ 
   0 &= D_{[\lambda}H_{\mu\nu\rho]}+  \frac{1}{4}\varepsilon_{\mu\nu\lambda\rho\sigma\tau}[Y^\sigma,X^I,D^\tau X^I]  + \frac{i}{8}\varepsilon_{\mu\nu\lambda\rho\sigma\tau}[Y^\sigma,\bar\Psi,\Gamma^{\tau}\Psi] \nonumber \\
 0&= \Gamma^\rho D_\rho\Psi + \Gamma_\rho\Gamma^I[Y^\rho,X^I,\Psi]  \ret
  0&= \tilde F_{\mu\nu}(\cdot) -[Y^\lambda,H_{\mu\nu\lambda},\ \cdot\ ] \ .
 \end{align}

Let's not worry about supersymmetry for now and look for  a lagrangian that reproduces these equations of motion. We will assume that all the constraints (\ref{constraints}) are imposed.  Months of trial and error lead to the following lagrangian\footnote{More precisely we should think of these as a family of lagrangians parameterised by the choice of $Y$ whose interacting part is five-dimensional.}
 \begin{align}\label{NAaction}
S =   \int d^6x\ \Big[ &\frac14\langle {\cal  D} B  \wedge  \star  {\cal  D} B \rangle
 +\frac{1}{6}\langle {\cal D}B  \wedge  DB \rangle  + \frac14\langle  \tilde  {  W}(Y)\wedge  \star  \tilde  { W}(Y)\rangle \ret 
  & -\langle H  \wedge  ({\cal  D} B -\tilde{  W}(Y) ) \rangle  -\frac12\langle ({\cal  D}B-\star {\cal  D}B  )\wedge \tilde { W}(Y) \rangle 
  +  (\tilde F \wedge   \tilde  { W} ) 
 \ret &  - \frac12 \langle  D_\mu X^I   D^\mu X^I \rangle -\frac14\langle [Y^\mu, X^I,X^J] [Y_\mu, X^I,X^J]\rangle  \ret &+\frac {i}{2}\langle  \bar\Psi  \Gamma^\mu D_\mu \Psi\rangle  + \frac {i}{2}\langle  \bar\Psi   \Gamma_\mu\Gamma^I [Y^\mu,X^I,\Psi]\rangle   \Big]\ ,
\end{align} 
 where 
\begin{align}
 \tilde  W (Y) &=  \frac{1}{3! }  W_{\mu\nu\lambda\rho}(Y^\rho)dx^\mu\wedge dx^\nu\wedge dx^\lambda\ ,
\end{align}
and we introduced the modified connection ${\cal D}_\mu  = { \partial}_\mu   - \tilde {\cal A}_\mu(\cdot )$ with
\begin{equation}
\tilde  {\cal A}_\mu (\cdot )  = \tilde A_{\mu }(\cdot )   - \frac12[B_{\mu\nu}, Y^\nu,\ \cdot\ ]\ .
\end{equation} 
 This derivative  has the effect of alleviating  the $D^2\ne 0$ problem that we mentioned above. In particular it enables the $\langle {\cal D}B\wedge DB\rangle$ term which vanishes if $\tilde {\cal A} = \tilde A$.

For $Y^\mu=0$ we obtain the flat-gauged theory above but for $Y^\mu\ne 0$ 
 the  Lagrange multiplier $\tilde W(Y)$ has  led to a source term for $H$ of the form
\begin{equation}
R = -\frac12\big(  \tilde W(Y)-\star \tilde  W(Y)\big)\ .
\end{equation} 
However we need to worry about the self-dual part of $\tilde W(Y)$. Without coupling this to something the equations of motion will be over constrained.  To this end we have  included a coupling of the self-dual part of   $\tilde W(Y)$  to the anti-self-dual part of ${\cal D}B$. This also can be accommodated by   a shift  $H\to H - \frac 12 \tilde W(Y) - \star\frac 12 \tilde W(Y)$.

Let us look at the    equations of motion. This action immediately reproduces the correct $X^I$ and $\Psi$ equations of (\ref{eomfixed}). The $H$ equation of motion implies that
\begin{equation}\label{Heq}
{\cal D}B - \tilde { W}(Y) = \star \big({\cal D}B - \tilde {  W}(Y)\big) \ ,
\end{equation}
whereas the $\tilde W$ equation of motion implies
\begin{align}
\tilde F_{\mu\nu}(\cdot) =[Y^\lambda,H_{\mu\nu\lambda},\ \cdot\ ] - \frac 32 [Y^\lambda,({\cal D}B-\star {\cal D}B)_{\mu\nu\lambda},\ \cdot\ ] - \frac 12 [Y^\lambda, \star \tilde {  W}(Y)_{\mu\nu\lambda},\ \cdot\ ] \ .
\end{align}
Putting these together we find 
\begin{align}
\tilde F_{\mu\nu}(\cdot) &=[Y^\lambda,H_{\mu\nu\lambda},\ \cdot\ ] - \frac 12 [Y^\lambda, \tilde { W}(Y)_{\mu\nu\lambda},\ \cdot\ ] \ret
& =[Y^\lambda,H_{\mu\nu\lambda},\ \cdot\ ]  \ ,
\end{align}
where in the last line we have used the constraint $[Y^\mu,Y^\nu,\ \cdot \ ]=0$ along with the fundamental identity. Thus we find agreement with (\ref{eomfixed}). 

Next let us examine the $\tilde A$ equation of motion
\begin{align}
 D^\mu (\star \tilde W)_{\mu\nu} = &-\Big[B^{\sigma\tau} , \Big(\frac12 {\cal D}B+ H  -\frac12  \tilde {  W}(Y)-\frac12 \star \tilde {  W}(Y)\Big)_{\nu\sigma\tau},\ \cdot\ \Big]\ret
 &+\frac{1}{24}\varepsilon_{\nu\mu\rho\alpha\beta\gamma}[B^{\mu\rho},[B^{\alpha\beta},B^{\gamma\sigma},Y_\sigma],\ \cdot \  ]\ret
& -[X^I,D_\nu X^I, \ \cdot\ ] - \frac{i}{2}[\bar\Psi,\Gamma_\nu\Psi,\ \cdot\ ]\ ,
\end{align}  
where $\star \tilde W$ is the  two-form Hodge dual of $\tilde W$ and the second line comes from the $\langle {\cal D}B\wedge DB\rangle$ term. 
Finally the $B$ equation of motion is
\begin{align}\label{Beq}
 0= &-\frac12 {\cal D}^\mu\Big(H + \frac12  {\cal D}B - \frac 12 \tilde W(Y) - \frac12\star\tilde W(Y)\Big)_{\mu\nu\lambda}\ret
&-\frac 12 \Big[B^{\sigma\tau},Y_{[\nu},\Big(\frac12 {\cal D}B  + H  - \frac12 \tilde W(Y)-\frac12\star \tilde W(Y) \Big)_{\lambda]\sigma\tau}\Big]\ret
&-\frac{1}{16}\varepsilon_{\nu\lambda\mu\alpha\beta\gamma}[B^{\mu\rho},Y_\rho,D^\alpha B^{\beta\gamma}]\ ,
\end{align} 
where the second line arises from the non-trivial dependence of ${\cal D}$ on $B$ and the third line from the $\langle {\cal D}B\wedge DB\rangle$ term. 
Remarkably,
putting all these equations together we simply find  
\begin{align}\label{DH}
0=& D_{[\lambda}H_{\mu\nu \rho]}   
+\frac14\varepsilon_{\mu\nu\lambda\rho\sigma\tau}  [X^I,  D^\tau X^I,  Y^\sigma] + \frac{i}{8}\varepsilon_{\mu\nu\lambda\rho\sigma\tau}[\bar\Psi,  \Gamma^\tau\Psi,   Y^\sigma ] \ ,
\end{align} 
which exactly reproduces the $H$ equation in (\ref{eomfixed}). In particular the field $B$ has decoupled in the sense that it does not appear in the equations of motion for the $X^I,H$ and $\Psi$ fields.

 Last but not least one can check that the action (\ref{NAaction}) is invariant under  the supersymmetry transformations
\begin{align}\label{NAsusy}
  \delta X^I &= i\bar\epsilon \Gamma^I\Psi\nonumber\\
  \delta B_{\mu\nu} &  = -i\bar\epsilon\Gamma_{\mu\nu}\Psi\ret
  \delta \Psi  &= \Gamma^\mu\Gamma^I D_\mu X^I\epsilon + \frac{1}{2\cdot 3!}H_{\mu\nu\lambda}\Gamma^{\mu\nu\lambda}\epsilon -\frac{1}{2}\Gamma_\mu\Gamma^{IJ}[Y^\mu,X^I,X^J]\epsilon \nonumber\\
  \delta H_{\mu\nu\lambda}  &= \frac{3}{2}i\bar\epsilon \Gamma_{[\mu\nu}D_{\lambda]}\Psi  +  \frac{3i }{{2}\cdot 3!}\varepsilon_{\mu\nu\lambda\rho\sigma\tau}\bar\epsilon\Gamma^{\rho\sigma}D^{\tau}\Psi - i \bar\epsilon\Gamma_\rho\Gamma_{\mu\nu\lambda}\Gamma^I[Y^\rho,X^I,\Psi]
  \nonumber\\ \delta \tilde A_\mu(\ \cdot\ )  &= i\bar\epsilon\Gamma_{\mu\nu}[Y^\nu,\Psi,\ \cdot\ ]\ret
  \delta \tilde W_{\mu
  \nu\lambda\rho}  &=     {3 i} \bar\epsilon\Gamma_{[\mu\nu}[B_{\lambda\rho] },\Psi, \ \cdot\ ] +i\bar\epsilon\Gamma_{\mu\nu\lambda\rho}\Gamma^I[X^I, \Psi,\ \cdot\ ] \ .  
  \end{align}  
For $\delta X^I$, $\delta \tilde A$ and $\delta \Psi$ these  transformations agree with the on-shell supersymmetries  of \cite{Lambert:2010wm} but $\delta H$ differs as here it must be self-dual off-shell (but agrees when the fermions are on-shell). 
In addition since we have recovered the equations of motion  of \cite{Lambert:2010wm} it follows that the supercurrent 
 \begin{equation}
S^\mu=-2\pi i\langle D_\nu X^I ,\Gamma^\nu\Gamma^I\Gamma^\mu \Psi\rangle+\frac{\pi i}{3!} \langle H_{\rho\sigma\tau },\Gamma^{\rho\sigma\tau}\Gamma^\mu\Psi\rangle-\pi i\langle[Y_\nu,X^I,X^J],\Gamma^\nu\Gamma^{IJ}\Gamma^\mu\Psi\rangle\ ,  
\end{equation}
obtained in \cite{Lambert:2011gb} is conserved.

As before the action involves the  fields of the $(2,0)$ tensor multiplet plus an additional self-dual three-form $B$ and gauge field one-form $\tilde A$.  The naive extension of the abelian case,  $H_{(s)} = \frac 12(DB+\star DB)+H$, is no longer a supersymmetry singlet as $\delta \tilde A\ne 0$.  However for the supersymmetry transformations (\ref{NAsusy}) one finds that\begin{equation}
{\cal H}_{(s)} = \frac12 \big({\cal D}B - \tilde { W}(Y)\big)+ \frac12 \star \big({\cal D}B - \tilde {W}(Y)\big)  +H\ .
\end{equation}
satisfies $\delta {\cal H}_{(s)}=0$. Furthermore the gauge field 
$\tilde {\cal A}_{(s)} = 2\tilde {\cal A} -   \tilde A$ is also a supersymmetry singlet: $\delta \tilde {\cal A}_{(s)}=0$.

   \section{Conclusions}\label{sect: Conclusions}
    
In this paper we have applied the construction of \cite{Sen:2015nph,Sen:2019qit} to the   action of a $(2,0)$ tensor multiplet in six dimensions.   In particular in section 3 we constructed an action for the abelian, free, $(2,0)$ tensor multiplet by introducing an additional two-form field and identifying a certain linear combination of the resulting self-dual three forms as a supersymmetry singlet. We also discussed how one might introduce an external source for the self-dual three-form. In section 4 we constructed a non-abelian action for the interacting $(2,0)$ system of equations of \cite{Lambert:2010wm}. This lead to a family of Lagrangians, parameterized by a choice of a three-algebra valued vector $Y^\mu$ which have six-dimensional Lorentz covariance which are invariant under a $(2,0)$ supersymmetry. The appearance of a covariantly constant vector is reminiscent of the PST construction \cite{Pasti:1996vs}. However the interacting part of the lagrangian is constrained to only depend on the coordinates orthogonal to $Y^\mu$.

One of the main goals of this work was to explore the sorts of constructions and structures that might feed into a better understanding of the non-abelian $(2,0)$ theories, whether or not  a lagrangian of sufficient utility exists. We don't expect to be successful in constructing a lagrangian that unambiguously defines the $(2,0)$ theory but we do hope that our discussion could have some use. For example the appearance of the two derivatives $D$ and $\cal D$    are curious. Since $Y$ picks an isometric direction this kind of coupling of the $B$-field is also reminiscent of a local manifestation of the ideas presented in \cite{Ganor:2017rrz}. 

More generally perhaps there are several lagrangian  descriptions, each of which captures some aspects of the $(2,0)$ theory, and that we should learn how to somehow patch these together, like charts covering a manifold.  In particular the   $(2,0)$ systems discussed here are parameterised by a choice of $Y^\mu$ and fall into three categories depending on whether $Y^\mu$ is spacelike, timelike or null. In each of these cases a maximally supersymmetric lagrangian in five-dimensions  does exist, see \cite{Lambert:2010wm,Hull:2014cxa,Lambert:2018lgt} respectively. In this sense the main idea of the non-abelian section of this paper is to find a unifying  six-dimesional  lagrangian structure  for these. It could be insightful to reproduce those lagrangians from the one presented here.     
It would also be interesting to see if one  generalise  this action to include M2-branes as in \cite{Lambert:2016xbs}.

 Even for the abelian case it could be interesting to compactify  it on $S^1$ using the non-standard coupling to a background metric and hence  $S^1$ radius    that arises \cite{Sen:2015nph,Sen:2019qit}. This might provide an alternative perspective that can circumvent the argument of \cite{Witten:2009at}.
   
In addition we would like to comment that  although the flat gauged theory constructed in section \ref{sect: Flat} may not seem very profound  the vacuum moduli space of M2-branes can also be obtained in this way, starting from a free theory. Indeed  the ABJM model also includes a gauge field that is a supersymmetry singlet. In particular in three-dimensions $\tilde W$ is a one-form and for abelian gaugings  the Lagrange multiplier term $\tilde F\wedge \tilde W$ can be re-written as a difference of two Chern-Simons terms with opposite levels. In that case the flat-gauged theory arises  as the low energy effective action on the M2-brane vacuum moduli space and plays an important role in the eleven-dimensional spacetime interpretation.  So perhaps one can make more sense of $\tilde F\wedge \tilde W$ term in six dimensions, without necessarily understanding the full non-abelian theory.

\section*{Acknowledgements}

I would like to thank C. Papageorgakis and A. Sen for communications as well as the organisers and participants of the Workshops "Higher Structures in M-Theory"  in Durham and  "String and M-Theory: The New Geometry of the 21st Century" in Singapore. This work was supported in part by STFC grant grant ST/L000326/1.


\begin{thebibliography}{10}


\bibitem{Lambert:2010wm}
  N.~Lambert and C.~Papageorgakis,
  JHEP {\bf 1008} (2010) 083
  doi:10.1007/JHEP08(2010)083
  [arXiv:1007.2982 [hep-th]].

\bibitem{Witten:1996hc}
  E.~Witten,
  J.\ Geom.\ Phys.\  {\bf 22} (1997) 103
  doi:10.1016/S0393-0440(97)80160-X
  [hep-th/9610234].
  
\bibitem{Tachikawa:2011ch}
  Y.~Tachikawa,
  JHEP {\bf 1111} (2011) 123
  doi:10.1007/JHEP11(2011)123
  [arXiv:1110.0531 [hep-th]].
  
\bibitem{Witten:2009at}
  E.~Witten,
  arXiv:0905.2720 [hep-th].
  
\bibitem{Bekaert:1999dp}
  X.~Bekaert, M.~Henneaux and A.~Sevrin,
  Phys.\ Lett.\ B {\bf 468} (1999) 228
  doi:10.1016/S0370-2693(99)01239-3
  [hep-th/9909094].
  
\bibitem{Chang:2018xmx}
  C.~M.~Chang,
  arXiv:1810.04169 [hep-th].
  
\bibitem{textbook}
Any text book on quantum field theory.

\bibitem{Sen:2015nph}
  A.~Sen,
  JHEP {\bf 1607} (2016) 017
  doi:10.1007/JHEP07(2016)017
  [arXiv:1511.08220 [hep-th]].
  
\bibitem{Sen:2019qit}
  A.~Sen,
  arXiv:1903.12196 [hep-th].
  
\bibitem{Samtleben:2011fj}
  H.~Samtleben, E.~Sezgin and R.~Wimmer,
  JHEP {\bf 1112} (2011) 062
  doi:10.1007/JHEP12(2011)062
  [arXiv:1108.4060 [hep-th]].
  
\bibitem{Mason:2011nw}
  L.~J.~Mason, R.~A.~Reid-Edwards and A.~Taghavi-Chabert,
  J.\ Geom.\ Phys.\  {\bf 62} (2012) 2353
  doi:10.1016/j.geomphys.2012.08.001
  [arXiv:1111.2585 [hep-th]].
    
\bibitem{Saemann:2011nb}
  C.~Saemann and M.~Wolf,
  J.\ Math.\ Phys.\  {\bf 54} (2013) 013507
  doi:10.1063/1.4769410
  [arXiv:1111.2539 [hep-th]].
  
  
\bibitem{Jurco:2019bvp}
  B.~Jurčo, T.~Macrelli, L.~Raspollini, C.~Sämann and M.~Wolf,
  arXiv:1903.02887 [hep-th].

  
\bibitem{Saemann:2017rjm}
  C.~Saemann and L.~Schmidt,
  arXiv:1705.02353 [hep-th].

\bibitem{Saemann:2019dsl}
  C.~Saemann and L.~Schmidt,
  arXiv:1908.08086 [hep-th].
  
  
\bibitem{Mkrtchyan:2019opf}
  K.~Mkrtchyan,
  arXiv:1908.01789 [hep-th].
  
  
\bibitem{Papadopoulos:2008sk}
  G.~Papadopoulos,
  JHEP {\bf 0805} (2008) 054
  doi:10.1088/1126-6708/2008/05/054
  [arXiv:0804.2662 [hep-th]].
  
\bibitem{Gauntlett:2008uf}
  J.~P.~Gauntlett and J.~B.~Gutowski,
  JHEP {\bf 0806} (2008) 053
  doi:10.1088/1126-6708/2008/06/053
  [arXiv:0804.3078 [hep-th]].
  
\bibitem{Lambert:2011gb}
  N.~Lambert and P.~Richmond,
  JHEP {\bf 1202} (2012) 013
  doi:10.1007/JHEP02(2012)013
  [arXiv:1109.6454 [hep-th]].
  
\bibitem{Pasti:1996vs}
  P.~Pasti, D.~P.~Sorokin and M.~Tonin,
  Phys.\ Rev.\ D {\bf 55} (1997) 6292
  doi:10.1103/PhysRevD.55.6292
  [hep-th/9611100].
  
\bibitem{Ganor:2017rrz}
  O.~J.~Ganor,
  Phys.\ Rev.\ D {\bf 97} (2018) no.4,  041901
  doi:10.1103/PhysRevD.97.041901
  [arXiv:1710.06880 [hep-th]].
 
\bibitem{Hull:2014cxa}
  C.~M.~Hull and N.~Lambert,
  JHEP {\bf 1406} (2014) 016
  doi:10.1007/JHEP06(2014)016
  [arXiv:1403.4532 [hep-th]].
   
\bibitem{Lambert:2018lgt}
  N.~Lambert and M.~Owen,
  JHEP {\bf 1810} (2018) 133
  doi:10.1007/JHEP10(2018)133
  [arXiv:1808.02948 [hep-th]].
  
   
\bibitem{Lambert:2016xbs}
  N.~Lambert and D.~Sacco,
  JHEP {\bf 1609} (2016) 107
  doi:10.1007/JHEP09(2016)107
  [arXiv:1608.04748 [hep-th]].

  
\end{thebibliography}
\end{document}